\documentclass[preprint2]{aastex}

\let\oldbibliography\thebibliography
\renewcommand{\thebibliography}[1]{%
  \oldbibliography{#1}%
  \setlength{\itemsep}{0pt}%
}

\shorttitle{Parque Astron\'omico de Atacama}
\shortauthors{Bustos et al.}

\begin{document}

\title{Parque Astron\'omico de Atacama: An Ideal Site for Millimeter, Submillimeter, and Mid-Infrared Astronomy}


\author{R. Bustos\altaffilmark{1}, M. Rubio\altaffilmark{2,3}, A. Ot\'arola\altaffilmark{4,5}, and N. Nagar\altaffilmark{6}} 


\altaffiltext{1}{Facultad de Ingenier\'ia, Universidad Cat\'olica de la Sant\'isima Concepci\'on, Alonso de Ribera 2850, Concepci\'on, Chile; rbustos@ucsc.cl.} 
\altaffiltext{2}{Departamento de Astronom\'ia, Universidad de Chile, Casilla 36-D, Santiago, Chile}
\altaffiltext{3}{Programa Astronom\'ia, CONICYT, Moneda 1375, Piso 7, Santiago, Chile}
\altaffiltext{4}{TMT Observatory, 1111 South Arroyo Parkway, Pasadena, CA 91105, USA}
\altaffiltext{5}{Atmospheric Sciences, University of Arizona, 1118 E 4th St, Tucson, Arizona 85721, USA}
\altaffiltext{6}{Departamento de Astronom\'ia, Universidad de Concepci\'on, Casilla 160-C, Concepci\'on, Chile}


\begin{abstract}
The area of Chajnantor, at more than 5000\,m altitude in northern Chile, offers unique atmospheric and operational conditions which arguably make it the best site in the world for millimeter, submillimeter, and mid-infrared observatories. Long-term monitoring of the precipitable water vapor (PWV) column on the Chajnantor plateau has shown its extreme dryness with annual median values of 1.1\,mm. Simultaneous measurements of PWV on the Chajnantor plateau (5050\,m) and on Cerro Chajnantor (5612\,m) show that the latter is around 36\% lower under normal atmospheric conditions and up to 80\% lower than the plateau in the presence of temperature inversion layers. Recently, the Government of Chile has consolidated the creation of the Parque Astron\'omico de Atacama (Atacama Astronomical Park), an initiative of the Chilean Commission for Science and Technology (CONICYT). This new park offers an opportunity for national and international projects to settle within its boundaries, gain access to an extremely dry site that is suitable for observations over a broad spectral range, especially in the millimeter to mid-infrared wavelengths, and benefit from operational and logistical support within a secure legal framework.
\end{abstract}




\section{Introduction}

Astronomical objects emit electromagnetic waves over a wide range of frequencies, or wavelengths. Our atmosphere, mainly due to the absorption of energy by atmospheric water vapor, partially or totally blocks sections of the electromagnetic spectrum that carry information about the cosmos. To mitigate this natural blocking, which is suffered by any ground-based astronomical observatory, a high altitude (to reduce the atmospheric column) and/or extremely dry (low water vapor column) site is desirable. The rare combination of both conditions creates an ideal site for millimeter, submillimeter, and mid-infrared observations from the ground. The Chajnantor area in Chile (23\degr\,1.4\arcmin\,S, 67\degr\,45.2\arcmin\,W), at over 5000\,m above sea level on the western side of the Andes mountains and about 50\,km east of San Pedro de Atacama, provides these particular conditions.

\begin{figure*}[ht]
\centering
\epsscale{1.8}
\plotone{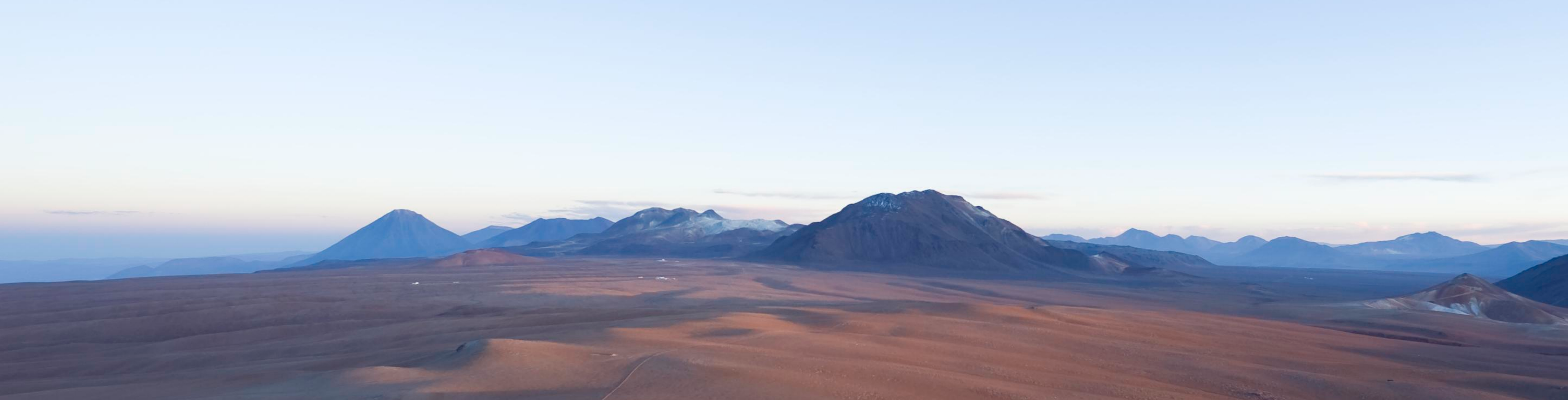}
\caption{Panoramic view of the Chajnantor area taken from Cerro Honar in 2007. To the \textit{left}, Volc\'an Licancabur (5920\,m). To the \textit{center-right}, Cerro Chajnantor (5640\,m). Cerro Toco (5604\,m) is in between. The \textit{white dots} on Llano de Chajnantor are the CBI and ALMA (under construction). To the \textit{right}, the Pampa La Bola plateau (4865\,m). Photo credit: ALMA (ESO, NAOJ, NRAO).\label{fig-1}}
\end{figure*}

Since 1997, several state-of-the-art instruments, covering different wavelength regimes, have settled at various sites near the Chajnantor area. Some have settled on the Chajnantor plateau (5050\,m): the Cosmic Background Imager (CBI; \cite{padin02}), the Atacama Pathfinder Experiment (APEX; \cite{gusten06}), the Q/U Imaging Experiment (QUIET; \cite{quiet13}), and the Atacama Large Millimeter Array (ALMA; \cite{wootten09}). Others, over various time spans, have settled on a plateau (5200\,m) on the nearby Cerro Toco (5604\,m): the Mobile Anisotropy Telescope (MAT; \cite{torbet99}), the Q-band Mobile Anisotropy Probe (QMAP; \cite{miller02}), the Millimeter-wave Interferometer (MINT; \cite{fowler05}), the Atacama Cosmology Telescope (ACT; \cite{swetz11}), the Radiative Heating in Underexplored Bands Campaign (RHUBC II; \cite{turner12}), Polarbear \citep{kermish12}, and the Atacama B-Mode Search (ABS; \cite{kusaka14}). The plateau (5525\,m) on Volc\'an Sairecabur hosted the Receiver Lab Telescope (RLT; \cite{marrone04}). On the nearby Pampa La Bola (4865\,m): the Atacama Submillimeter Telescope (ASTE; \cite{kohno05}) and the NANTEN2 Observatory \citep{kawamura05}. On Cerro Chajnantor (5640\,m),  a 1\,m telescope (mini-TAO) was installed as a pathfinder for a future 6\,m mid-infrared telescope \citep{sako08}. New projects such as the Cosmology Large Angular Scale Surveyor (CLASS), High Altitude Terahertz Solar Telescope (HATS), CCAT, Tokyo Atacama Observatory (TAO), and Chilean Medidor Aut\'onomo de Radio Interferencia (MARI) are planning to use the Chajnantor area in the near future. 

Based on transparency measurements \citep{radford98} and the number of installed and planned telescopes, the Chajnantor area has become a leading global site for millimeter and submillimeter astronomy. Similarly, based on the first observations by mini-TAO, the area has proven to be excellent for observations in the mid-infrared regime \citep{sako08, motohara08}.

Figure~\ref{fig-1} shows a panoramic view of the Chajnantor plateau and the nearby mountains, looking north from Cerro Honar in 2007. The summit of Cerro Chajnantor is almost 600\,m higher than the plateau, and thus has a significantly lower atmospheric moisture column. Both Cerro Toco and Cerro Chajnantor, as well as their immediate surroundings, are inside the boundaries of the Parque Astron\'omico de Atacama.

The extremely dry conditions of the Atacama Desert result from a combination of several effects, of which the primary is its subtropical location, which is right under the descending branch of the Hadley cell of dry and warm air, whose subsidence maintains the southeast Pacific anticyclone. This high pressure keeps the middle and lower troposphere warm and provides conditions that explain both the aridity and dryness of this subtropical region \citep{garreaud11}. The subsiding air that creates this high surface pressure is also responsible for the strong temperature inversion layer that gives origin to the southeast pacific maritime cloud deck \citep{munoz10}. The coastal mountain range on the west side of Chile, at over 2000\,m altitude, combined with the temperature inversion layer, block the moist air flow to the interior of this region. To the east side, the Andes mountain range blocks the humidity flow from the Amazon basin, except in Austral summer, the so-called Altiplanic Winter \citep{bustos01}. These effects, in combination with high elevation plateaus on the western side of the Andes, produce ideal sites to scientifically exploit the millimeter, submillimeter, and mid-infrared spectral windows. 

\begin{figure*}[ht]
\centering
\epsscale{1.5}
\plotone{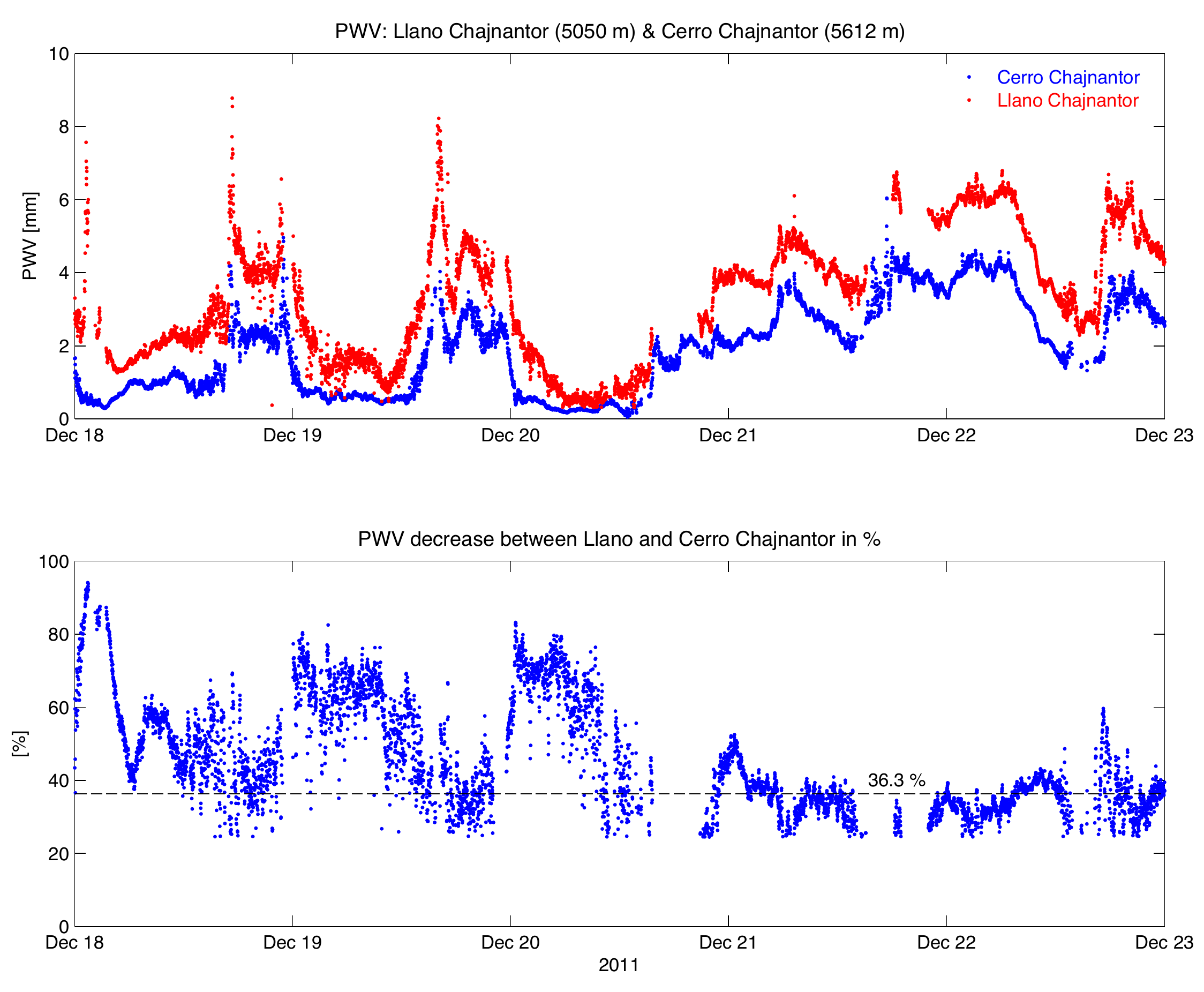}
\caption{\textit{Top}: A comparison of the 1-minute averages of the simultaneous PWV measurements at Llano de Chajnantor (\textit{red}) and Cerro Chajnantor (\textit{blue}) over a 5 day run. \textit{Bottom}: The drop in PWV in percent between Llano and Cerro Chajnantor. The \textit{dotted line} shows the 36.3\% decrease in PWV.\label{fig-2}}
\end{figure*}

Various methods are used to characterize the atmospheric transmission at a given site. Most are based on passive radiometry to monitor the sky brightness within a given spectral band. On Chajnantor, the atmospheric transparency at 225\,GHz was monitored as part of the ALMA site-testing effort using a 225\,GHz tipping radiometer \citep{mckinnon88}. The Harvard-Smithsonian Center for Astrophysics \citep{paine00} and the Nobeyama Radio Observatory \citep{mat99} monitored the sky transparency using a Fourier transform spectrometer from 200\,GHz up to 3.5\,THz and 1.5\,THz, respectively. Another method to characterize the atmospheric transmission is to combine measurements of the integrated water vapor of the atmospheric column with a physical model of the atmospheric absorption spectrum as a function of atmospheric moisture. This method was followed by the ALMA site-testing team using two passive radiometers to monitor the intensity of the 183\,GHz water vapor emission line \citep{delgado99}. Considering that water vapor pressure is limited by temperature through the Clausius-Clayperon equation, the precipitable water vapor (PWV) content at any given height is modeled with an exponential decay in altitude at a rate given by the scale height $h_{0}$ \citep{otarola10}, as seen in equation~\ref{eq-1}. 
 
\begin{equation}\label{eq-1}
PWV(h)=PWV_{0}\cdot e^{-\frac{h}{h_{0}}}
\end{equation}

Here, $PWV_{0}$ (in millimeters) is the corresponding PWV at the surface level, and the scale height $h_{0}$ (in meters) corresponds to the e-folding altitude in the exponential distribution of water vapor with altitude. Deviations from the model of equation~\ref{eq-1} are due to horizontal advection of moisture at varying altitudes in the atmosphere and/or changes in water vapor density mainly due to the presence of temperature inversion layers in the troposphere \citep{roosen77,otarola10}.

For the Chajnantor area, \citet{giovanelli01} determined a median $h_{0}$ of 1135\,m from balloon measurements and \citet{bustos00} estimated $h_{0} \sim 1500$\,m from a reanalysis of historical data. Neither of these methods measured $h_{0}$ continuously. For $h_{0}=1135$\,m (1500\,m), the expected drop in PWV between Llano de Chajnantor (5050\,m) and Cerro Chajnantor (5612\,m) is 39.1\% (31.2\%). Here, we present preliminary results on simultaneous PWV measurements, directly obtaining the difference in PWV between both sites.

\section{PWV observations: Llano versus Cerro Chajnantor}

In this experiment, we used two radiometers that measured the intensity of the 183\,GHz water vapor line. One of these was originally built and used for the ALMA site-testing campaign \citep{delgado99}. This unit was refurbished by the Laboratory of Radio Astronomy at Universidad de Concepci\'on in 2009 with the installation of a new backend, replacement of the tipping mirror motors, and the implementation of new software for observing, control, and communications \citep{bustos11}. The second 183\,GHz radiometer used in the experiment is the one that is mounted in the cabin of the APEX radio telescope located on Llano de Chajnantor, whose data is available on the web\footnote{http://www.apex-telescope.org}.

The refurbished radiometer was first installed next to APEX for a 10-month, side-by-side calibration test in January 2011. This test showed that both PWV time-series compared well to a level of $\pm$0.14\,mm in 1-minute samples. At the end of 2011, the refurbished radiometer was relocated to the CCAT site weather monitoring station, slightly below the summit of Cerro Chajnantor, starting PWV data acquisition on 2011 December 17. From the Llano, the horizontal distance of these radiometers is 3\,km in the norhteast direction and the altitude difference is 562\,m. A snowstorm that started on 2011 December 22 produced a power failure on the solar panels of the CCAT station that was not resolved until October 2012. Hence, only a week of PWV data was obtained in this first campaign. 

\begin{figure}[h]
\epsscale{1}
\plotone{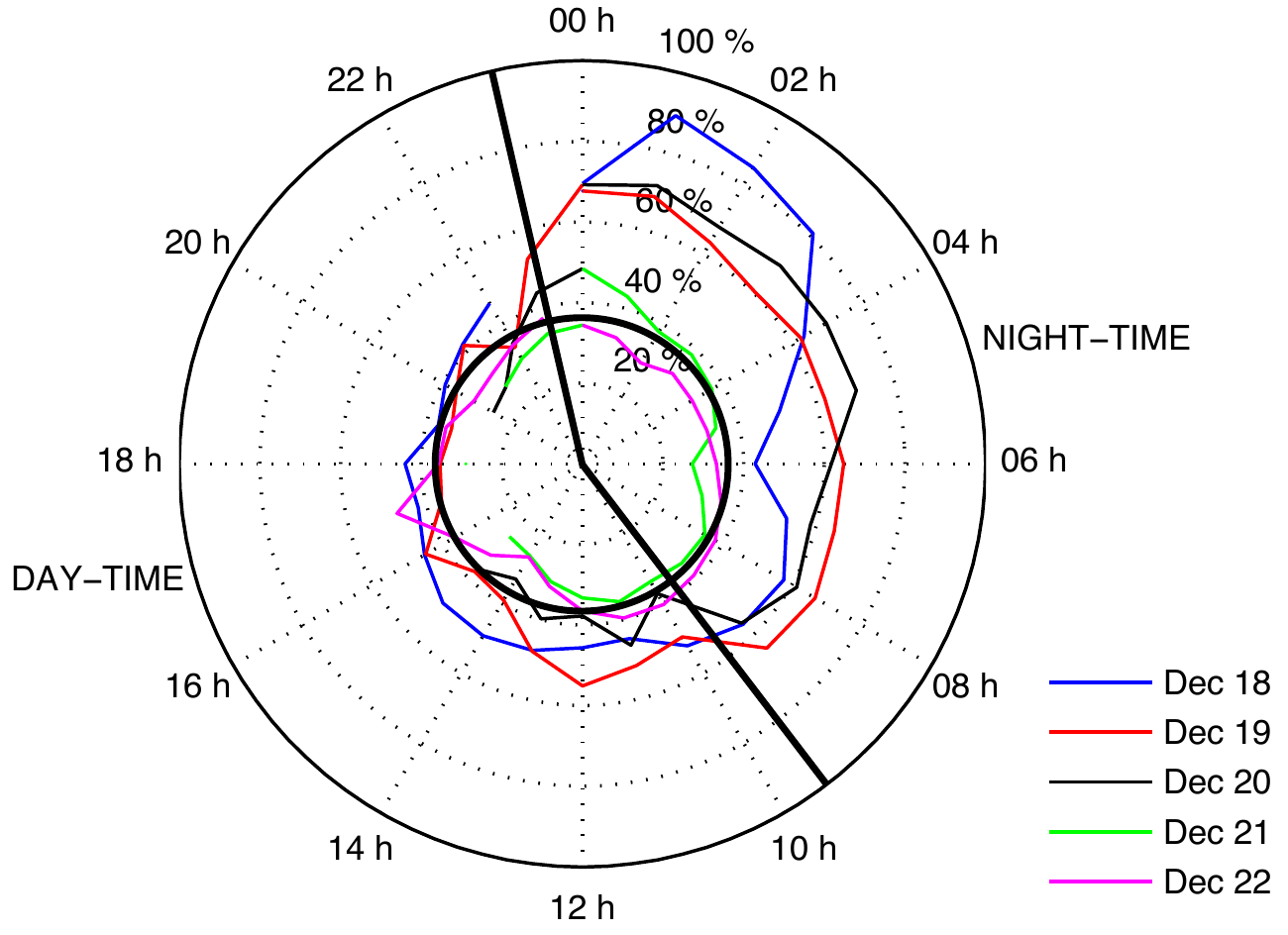}
\caption{24 h polar plot of the drop in PWV between the Llano and Cerro Chajnantor for each of the 5 days. Nighttime (23:00-9:30 UTC) and daytime are separated by the \textit{solid black line}. The \textit{black circle} is the 36.3\% level. \label{fig-3}}
\end{figure}

Simultaneous PWV observations on Llano de Chajnantor (APEX site) and Cerro Chajnantor (CCAT site) from 2011 December 18-23 are presented in Figure~\ref{fig-2}. This figure shows that under normal atmospheric conditions, without the presence of inversion layers between the two sites, the drop in PWV from the Llano to Cerro Chajnantor is around 36.3\% (\textit{dotted line}). This $\sim$36.3\% difference is more often seen in the afternoons, since the atmosphere is generally well-mixed and no significant inversion layers are expected. On the nights of December 18, 19, and 20, the extra difference in PWV (up to 90\%  on Dec. 18 and up to 70\%  on the other two nights) is likely explained by the formation of an inversion layer between the Llano and Cerro Chajnantor. From December 21 to 23, no evidence for a significant inversion layer is seen. This could be explained by the relatively high amount of PWV, which is a usual occurrance previous to snowstorms such as the one on December 22 that produced the power failure. 

To show that inversion layers between Llano and Cerro Chajnantor generally occur during night-time, Figure~\ref{fig-3} separates nighttime and daytime on a 24 h polar plot. Nighttime is considered to be between 23:00 and 9:30 UTC (8 p.m. to 6:30 a.m., local time). For December 18 to 20, the drop in PWV is significantly larger at nighttime compared to daytime.

\begin{figure*}[t]
\centering
\epsscale{1.6}
\plotone{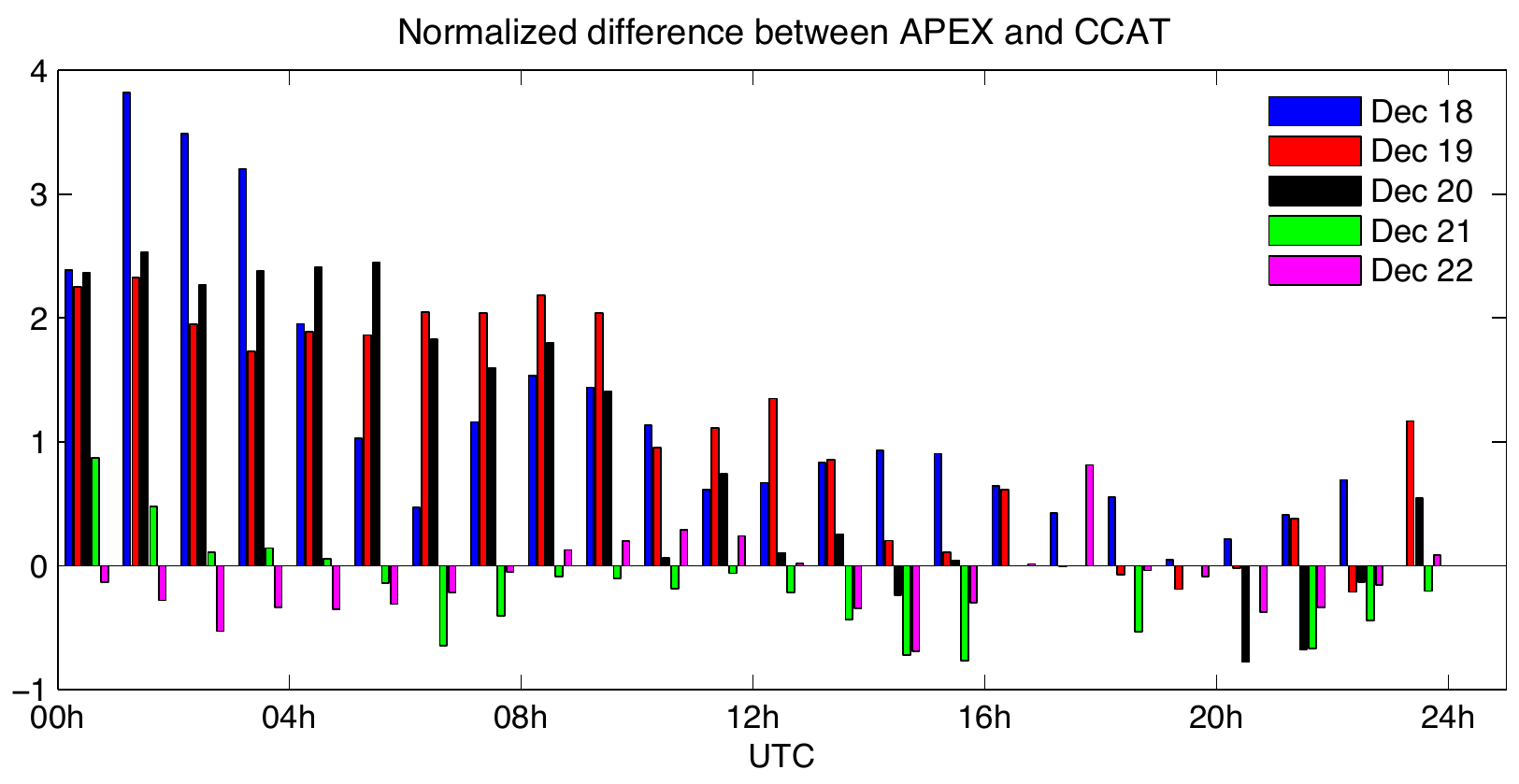}
\caption{Normalized PWV difference between Llano and Cerro Chajnantor, averaged over 1 h for each of the five days. The zero level corresponds to a PWV difference of 36.3\%. The standard deviation of all data together is 13.9\%.\label{fig-4}}
\end{figure*}

To determine the contribution and strength of the inversion layers, the drop in PWV was normalized to the 36.3\% level with a standard deviation $\sigma$ of 13.9\%. Figure~\ref{fig-4} shows the normalized drop in PWV between both sites for each hour and each day. As mentioned in Figure~\ref{fig-3}, the differences are stronger and more significant for the first three nights. During all five afternoons and in the last two days, the drop in PWV is within the 1-$\sigma$ level. 

Given the $\pm$0.14\,mm error in 1-minute samples we obtained between both radiometers in the side-by-side calibration test at the APEX site, we expect a statistical error in 1 h of $\pm$0.018\,mm. Thus, for a PWV of 1.0\,mm, the difference in PWV is expected to have a measurement error of 1.8\%, significantly lower than the 1-$\sigma$ level of 13.9\% seen in Figure~\ref{fig-4}. More details on the calibration campaign and a larger dataset will be presented in a following paper.

\section{Worldwide site comparison}

\cite{radford11} have compared the atmospheric transparency conditions at several important sites worldwide: Mauna Kea, Hawaii (4070\,m); the South Pole (2800\,m); Llano de Chajnantor (5050\,m); and Cerro Chajnantor (5612\,m). Cumulative distributions through several years were obtained by measuring the 350\,$\mu$m and 225\,GHz zenith optical depth at these four sites. They have shown that Cerro Chajnantor, Llano de Chajnantor, and the South Pole are usually drier than Mauna Kea. If we consider the superb observing conditions when 350\,$\mu$m zenith optical depths are below 1.1, the best sites are arguably found in Chile. In this case, Cerro Chajnantor presents 350\,$\mu$m zenith optical depths below 1.1 for 50\% of the time, which is twice that of Llano de Chajnantor and the South Pole. 

\begin{figure*}[t]
\epsscale{1.7}
\plotone{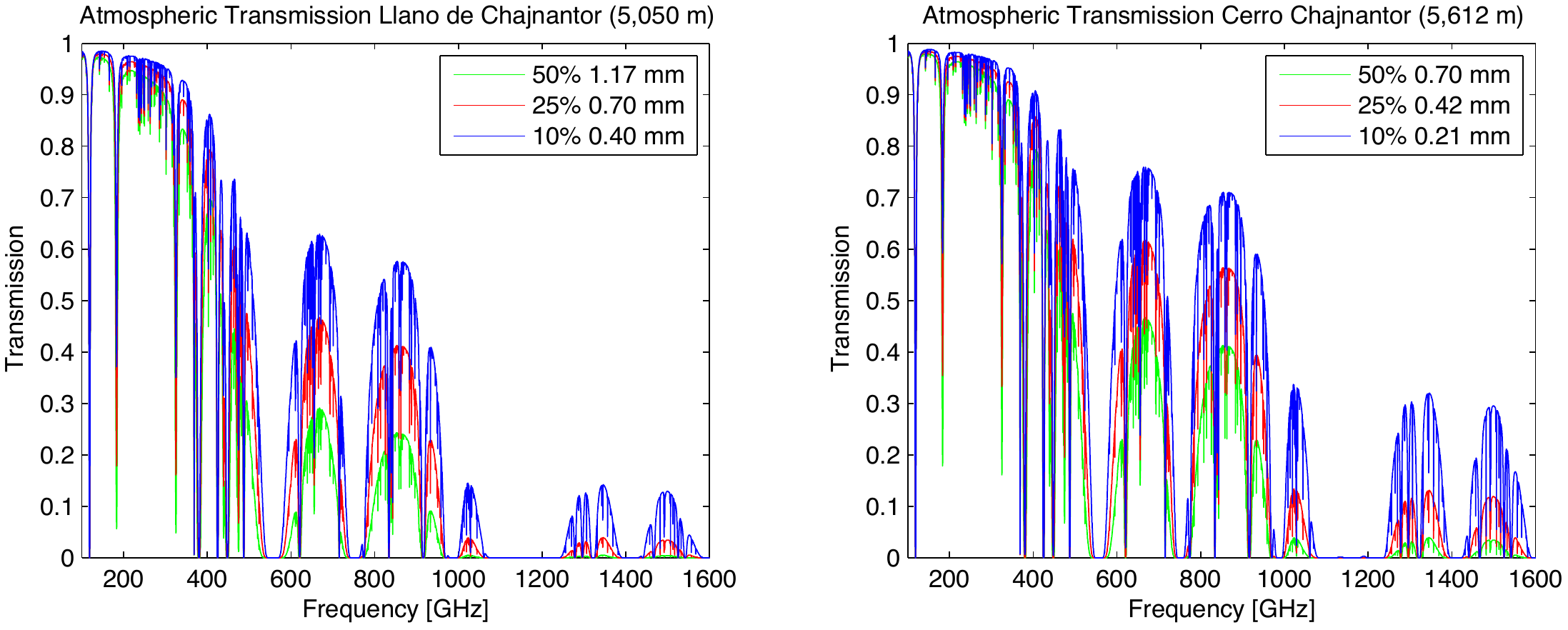}
\caption{Atmospheric transmission in the millimeter and submillimeter range for Llano de Chajnantor (\textit{left}) and Cerro Chajnantor (\textit{right}). Transmissions are derived from observed 350\,$\mu$m zenith optical depths and the ATM model \citep{pardo01}. The \textit{green}, \textit{red}, and \textit{blue lines} show the transmission for 50\%, 25\%, and 10\% of the time, respectively. The \textit{inset} lists the corresponding PWV in millimeters. Comparing \textit{left} and \textit{right panels} clearly shows the gain in atmospheric transmission by going to a higher site, especially at terahertz frequencies.\label{fig-5}}
\end{figure*}

Using the ATM model \citep{pardo01} and the 350\,$\mu$m zenith optical depth cumulative distributions obtained by \cite{radford11}, we compared the atmospheric transmissions at Llano de Chajnantor and Cerro Chajnantor (Figure~\ref{fig-5}). Cerro Chajnantor starts to show significant improvements in atmospheric transmission at frequencies above 600\,GHz. In the THz frequency regime, the atmospheric transmission on Cerro Chajnantor is twice that of Llano de Chajnantor. 

\begin{figure*}[ht]
\epsscale{1.7}
\plotone{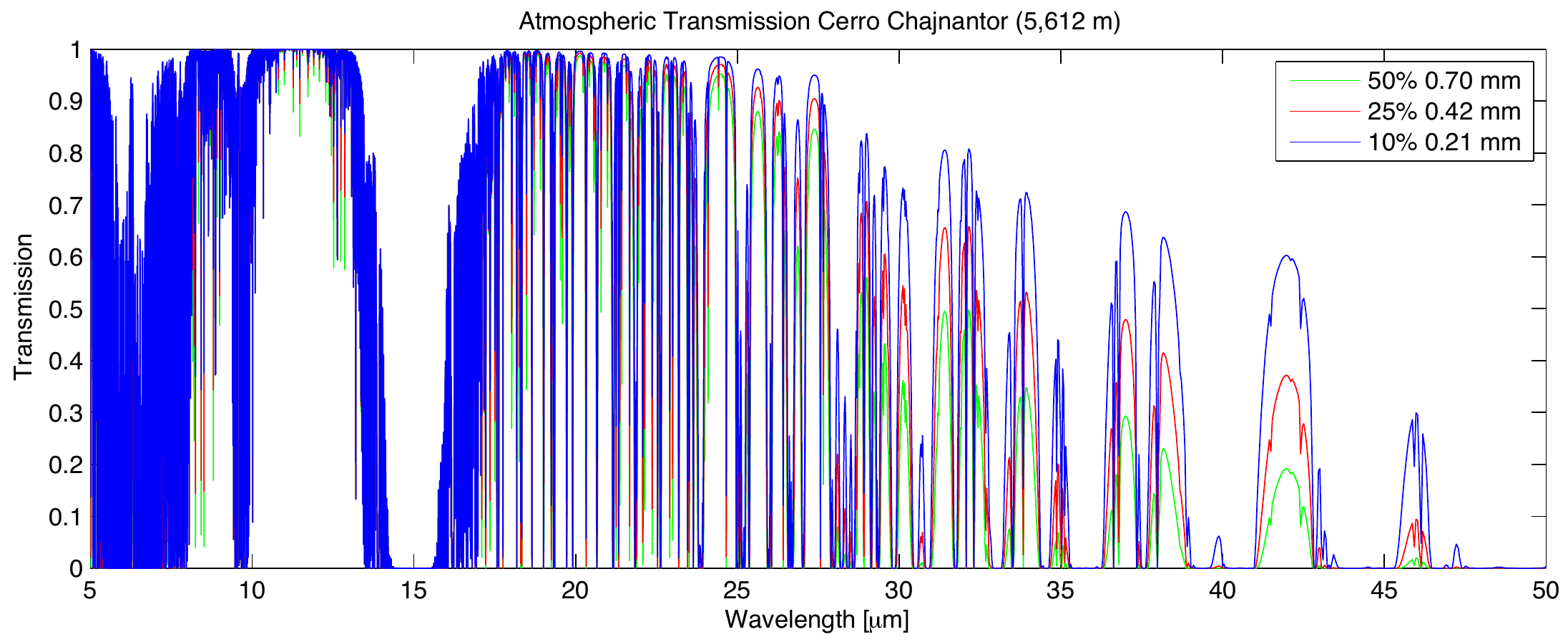}
\caption{Atmospheric transmission on Cerro Chajnantor in the mid-IR to far-IR bands. \textit{Green}, \textit{red}, and \textit{blue lines} are 50\%, 25\%, and 10\% of the time, respectively, with the corresponding PWV in millimeters.\label{fig-6}}
\end{figure*}

For the mid-infrared to far-infrared (mid-IR to far-IR) spectral regime, Cerro Chajnantor presents an atmospheric transparency of up to 40\% in the 36-38\,$\mu$m band \citep{sako08} and a median atmospheric seeing of 0.69\arcsec\, at 500\,nm \citep{motohara08}. In this work, we implemented a line-by-line, layer-by-layer radiative transfer model in Matlab to calculate the atmospheric transmission for Cerro Chajnantor (Figure~\ref{fig-6}). The model included a dry atmosphere composition (Table~\ref{table1}) and a wet contribution of 0.70\,mm (50\% of the time), 0.42\,mm (25\%), and 0.21\,mm (10\%) of PWV. The spectral information in the radiative transfer model comes from the HITRAN 2008 database at high spectral resolution \citep{rothman09}.

\begin{table}[t]
\begin{center}
\epsscale{0.5}
\caption{Dry atmosphere composition\label{table1}}
\begin{tabular}{lcl}
\tableline\tableline
Molecular species & Formula & Fraction per volume \\
\tableline
Nitrogen & N$_2$ & 0.7808\\
Oxygen & O$_2$ & 0.2095\\
Argon & Ar & 0.0093\\
Carbon Dioxide & CO$_2$ & 0.000387\\
Methane & CH$_4$ & 0.000001745\\
Carbon Monoxide & CO & 0.000000750\\
Nitrous Oxide & N$_2$O & 0.000000314\\
\tableline
\end{tabular}
\end{center}
\end{table}

Figure~\ref{fig-6} shows the resulting transmission spectra in the mid-IR to far-IR bands. The dryness of Cerro Chajnantor is such that for PWV conditions of 0.21\,mm (10\% of the time), the atmospheric transmission is higher than 95\% in the \textit{Q} (17-25\,$\mu$m) band, higher than 60\% in the \textit{Z} (28-40\,$\mu$m) band, and higher than 25\% in several observing windows beyond \textit{Z} band. Therefore, in extreme dry conditions (10\% of the time), it is expected that Cerro Chajnantor offers completely unique conditions for mid-IR to far-IR observations of cosmic sources from the ground. 

These results are in agreement with previous simultaneous measurements of 350\,$\mu$m zenith optical depths obtained in \cite{radford08} and \cite{radford11} and scale height values from balloon measurements in \cite{giovanelli01}.

Besides the excellent atmospheric conditions, other advantages of the Chajnantor area are its geographic location, accessibility, and logistics. From latitude 23\degr\,S, the southern sky and a good fraction of the northern sky is available for astronomical research. It has year-round accessibility via a paved road (the international route connecting Chile and Argentina). It is close to large cities (Calama, Antofagasta) with ports, airports, and infrastructure that can provide general services, communications, and energy by local companies. 

\section{Parque Astron\'omico de Atacama}

Given the unique and excellent transparency conditions for astronomical observations of the Chajnantor area, the Government of Chile has secured an area of 363.81\,km$^2$ of fiscal land for the exclusive use of scientific activities through the Parque Astron\'omico de Atacama\footnote{http://www.conicyt.cl/astronomia}. The Ministry of National Assets has given a long-term land concession (50 years, starting from 2013) to the National Commission for Research in Science and Technology (CONICYT) (Diario Oficial 40.688, 2013) and given CONICYT the responsibility for its administration. 

\begin{figure*}[ht]
\epsscale{1.6}
\plotone{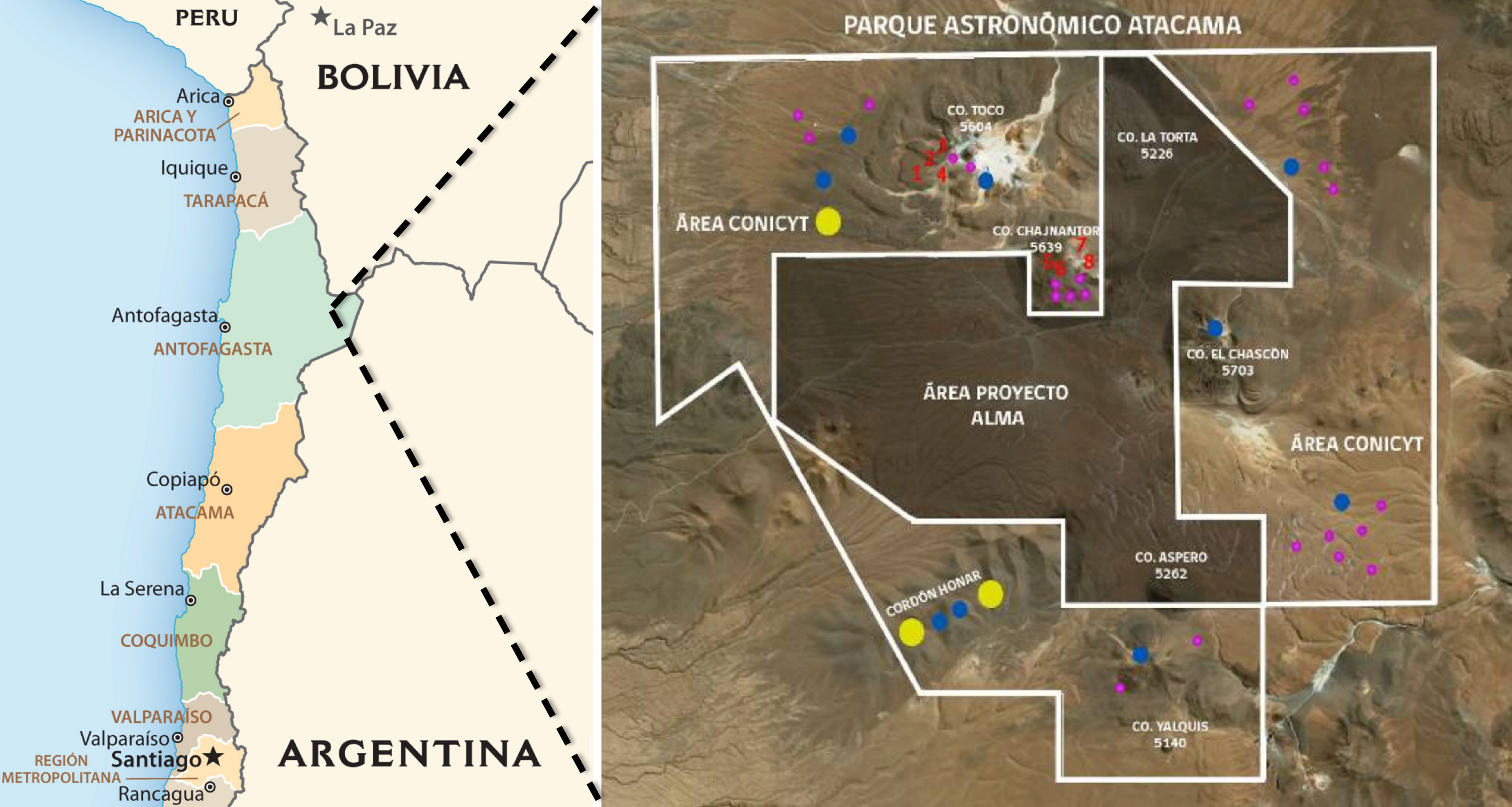}
\caption{\textit{Left}: The Chajnantor area in northern Chile, close to the border of Bolivia and Argentina. \textit{Right}: Close-up showing the area of the Parque Astron\'omico de Atacama and the ALMA concession (\textit{dark brown}). The \textit{colored dots} show potential sites where new telescopes can be installed in the park. \label{fig-7}}
\end{figure*}

In Figure~\ref{fig-7}, the left panel shows the location of the Parque Astron\'omico de Atacama in the II Region of Antofagasta in Chile, and the right panel shows its extension in the marked area. This area has been protected by the Chilean government for scientific research and astronomical observations since 1998 by Presidential Decree N\textsuperscript{o}\,185. Two short-term land concessions (5 years each) to CONICYT allowed the establishment of the CBI telescope in 1999, followed by ASTE, APEX, NANTEN-2, ACT, mini-TAO, and Polarbear. CONICYT transferred part of the protected area to the ALMA project in 2003.

The main purpose of this park is to promote the installation of new astronomical facilities, to help provide necessary infrastructure, and to coordinate the development of the area with the current and future observatories or telescopes. The park is open to host big international projects as well as national or international university telescopes that could be operated in person or remotely. Other types of projects which do not interfere with astronomical observations, i.e. meteorological monitoring, seismic stations, solar research, etc. are also welcome.

To achieve this purpose, in October 2013, the National Commission for Research in Science and Technology (CONICYT) created a foundation as the entity responsible for managing and developing the Parque Astron\'omico Atacama. The foundation will be responsible for preserving and protecting the area from mining, geothermic, or other industrial activities that could interfere with astronomical activities, as well as generating general norms and regulations for telescope installations, security procedures, coordination between different projects, and the provision of common services, among others. It will also maintain a relationship with local native communities and the local administration. 

The area has also been declared a radio quiet zone (no radio emissions above 31.3\,GHz) over a 30\,km radius and has a coordination zone of 120\,km radius to limit the granting of radio emission licenses by the Ministry of Telecommunications around the park. This radio protection adds to the existing light pollution protection by the Chilean government for the II, III and IV regions of the country issued by the Ministry of Environment (Decree No. 043, 2013) to preserve the dark skies over this large territory of the country.

\section{Conclusions}

Simultaneous measurements of PWV at the Llano de Chajnantor (5050\,m) and Cerro Chajnantor (5612\,m) were analyzed using 183\,GHz radiometers between 2011 December 18-23. We find that PWV decreases by $\sim$36\% under normal atmospheric conditions on Cerro Chajnantor as compared to Llano de Chajnantor. This decrease is consistent with a well developed boundary layer of at least 600\,m depth above the plateau. In the presence of inversion layers, the drop in PWV can be as high as 90\%, improving the observing conditions for the millimeter to mid-infrared astronomy. 

These measurements can constrain the scale height continuously, and therefore they provide a very useful tool for site characterization and radio telescope operations, and are an alternative method for determining the strength of inversion layers. They also help to illustrate the advantage of going to higher sites for astronomical observations at submillimeter and infrared wavelengths.

Technical aspects such as extremely low water vapor content, high transparency, and sky coverage as well as operational aspects such as accessibility, communications, and services makes the Chajnantor area unique for astronomical observatories.

The government of Chile, considering these aspects, has established the Parque Astron\'omico de Atacama administered by CONICYT to protect and develop this area, identified as the best site in the world for millimeter, submillimeter, and mid-infrared astronomy. 

\acknowledgments

We acknowledge support from Dr. Simon Radford and the CCAT project to install and operate the 183\,GHz radiometer. We thank support from the APEX staff and for the APEX PWV data. We also thank  Jos\'e Cort\'es and Hugo Pacheco for their help at the site and the CONICYT Astronomy Program. 

R.B. and N.N. work was supported by ALMA-CONICYT grants 31080022 and 31070015. R.B. acknowledges support from CONICYT through QUIMAL grant N\textsuperscript{o}\,130005. M.R. wishes to acknowledge support from CONICYT (Chile) through FONDECYT grant N\textsuperscript{o}\,1140839.

M.R. would like to thank the presidents of CONICYT, Dr. Eric Goles (2000-2006),  Mrs. Vivian Heyl (2006-2010), and Dr. Jos\'e Miguel Aguilera (2010-2013) for the permanent support of the Parque Astron\'omico Atacama project, and Mrs. Alicia Norambuena, Coordinator of the Parque Astron\'omico Atacama, for her strong commitment to the project and for ensuring the protection of the area from activities that could interfere with the astronomical projects. We would also like to thank the Ministry of National Assets for its support during the administrative processes of the land concessions.

We thank Dr. Martin Shepherd and an anonymous reviewer for useful comments that contributed to improve the reading of the manuscript.

\end{document}